# Wave solutions of Gilson-Pickering equation

**Karmina K. Ali*,1,2, Hemen Dutta[3], Resat Yilmazer[2] and Samad Noeiaghdam[4,5]**

[1]Faculty of Science, Department of Mathematics, University of Zakho, Iraq
[2]Faculty of Science, Department of Mathematics, Firat University, Elazig, Turkey
[3]Department of Mathematics, Gauhati University, Guwahati-781014, Assam, India
[4]Baikal School of BRICS, Irkutsk National Research, Technical University, Irkutsk, Russian Federation
[5]South Ural State University, Lenin prospect 76, Chelyabinsk, 454080, Russian Federation
karmina.ali@uoz.edu.krd, rstyilmazer@gmail.com, and samadnoeighdam@gmail.com

**Abstract:** In this work, we applied $(1/G')$-expansion method to produce the novel soliton solution of the Gilson–Pickering equation. This method is fundamental on homogenous balance procedure that gives the order of the estimating polynomial-type solution. Also it is based on the appreciate wave transform to reduce the governing equation. The solutions that we obtain are include of hyperbolic, complex and rational functions solutions. Finally, the results are graphically discussed.

**Keywords:** Gilson–Pickering equation, $(1/G')$-expansion method, trigonometric, complex, hyperbolic, and rational functions.

## 1. Introduction

We consider the one module of the nonlinear evolution equations (NLEEs), which is the fully nonlinear third-order partial differential equations (PDEs)

$$u_t - \epsilon u_{xxt} + 2ku_x - uu_{xxx} - \alpha uu_x - \beta u_x u_{xx} = 0, \tag{1}$$

where $\varepsilon, \alpha, \kappa$ and $\beta$ are non-zero real numbers. Which was introduced by Claire Gilson and Andrew Pickering in 1995 [1]. There is three special instances of Gilson-Pickering equation (GPE) have been seeming in the written material, these are, when $\alpha = -1$, $\epsilon = 1$, $k = 0.5$, and $\beta = 3$ then Eq. (1) is the Fornberg-Whitham (FW) equation [2], [3], [4], when $\alpha = 1$, $\epsilon = 0$, $k = 0$, and $\beta = 3$ then Eq. (1) is the Rosenau-Hyman (RH) equation [5], when $\alpha = -3$, $\epsilon = 1$, and $\beta = 2$ then Eq. (1) is the Fuchssteiner-Fokas-Camassa-Holm (CH) equation [6], [7]. There are many researchers investigated the Gilson-Pickering equation by different analytical methods of solving the governing equation such as Bernoulli sub-equation function method [8], not- a- knot meshless method [9], the first integral method [10], the $G'/G$ method [11].

The layout of this work is as follows: In section two, we present the $(1/G')$-expansion method, and how to use this method by explaining some necessary steps, the mentioned method was first time introduced by Asif yokus [12], [13], [14]. In section three we put through the ($1/G'$)-expansion method on the Gilson-Pickering equation and gives some family of equations that give us the different new solution of the (GPE), and graphed three dimensional, two dimensional and counter surfaces of the governing equation. In section four, we present a conclusion of our new work, at the end of this work we give some references.

## 2. Application of the $(1/G')$-expansion method:

Let us consider a nonlinear partial differential equation (NLPDE) as follows:

$$Q(u, u_t, u_x, u_{xx},...) = 0, \tag{2}$$

where $u = u(x,t)$ and its partial derivatives wherein the highest order derivatives and nonlinear terms are involved. The fundamental steps of the $(1/G')$-expansion method are as follows:

Step 1: let consider the traveling wave transformation as

$$u(x,t) = U(\xi)e^{i(\alpha x + \beta t)}, \quad \xi = k(x - vt), \tag{3}$$

where $v$ is the speed of the traveling wave, $k$ is the wave number, $\alpha$ and $\beta$ are nonzero constants, after some procedure, Eq. (3) reduce into the following nonlinear ordinary differential equation (NLODE):

$$Q(U', U'', U''',...) = 0, \tag{4}$$

where $Q$ is a polynomial in $U(\xi)$ and its derivatives, forasmuch $U'(\xi) = \frac{dU}{d\xi}$, on the other hand, the solution of the linear second order ODE is given below

$$G'' + \lambda G' + \mu = 0, \quad \text{where } G = G(\xi), \tag{5}$$

Step 2: Suppose the solution of Eq. (5) can be written in the form

$$U(\xi) = \sum_{i=0}^{m} A_i \left(\frac{1}{G'}\right)^i, \tag{6}$$

where $A_0, A_1, A_2, ..., A_m$ are constants and $m$ is the balancing term. To determine the values of $m$, we balance the linear term of the highest order to the highest order nonlinear term. Here is $(1/G')^i, i = 0,1,2,...,m$ matching the coefficients of the terms to zero, we obtain an algebraic equation system. This algebraic equation system is resolved manually or with the help of the computer package program. These solutions can be found in the solution function provided by (5) and are written in the form of (6).

## 3. Mathematical analysis:

In this section, the $(1/G')$-expansion method has been applied to the Gilson-pickering equation (e.g.,1) to find exact solutions, using the traveling wave transformation Eq. (1) as $u = U(\xi), \xi = x - ht,$ where $h$ is an arbitrary non-zero real number, obtaining the following nonlinear ordinary differential equation (NLODE):

$$(2k - h)U' + \epsilon h U''' - UU''' - \beta U'U'' - \alpha UU' = 0, \tag{7}$$

where $\epsilon, \beta, \alpha$ and $k$ are non-zero real numbers.

Integrating Eq. (7) once with respect to $\xi$ and assuming that the constant of integration to be zero, we obtain

$$(2k-h)U + (\epsilon h - U)U'' + \frac{1-\beta}{2}(U')^2 - \frac{\alpha}{2}U^2 = 0, \qquad (8)$$

we obtain $m=2$ by using the balancing homogenous principle. Using Eq. (6) together with $m=2$, we have:

$$U(\xi) = a_0 + a_1\left(\frac{1}{G'}\right) + a_2\left(\frac{1}{G'}\right)^2, \qquad (9)$$

where $a_0, a_1, a_2$ are constants. Putting Eq. (9) into Eq. (8), getting a system of a trigonometric function, solving this system by some computational program like as Matlab and Mathematica, gives new hyperbolic solutions, trigonometric solutions, and complex solutions, as follows:

Set 1:

$$a_0 = -\frac{2(h-2k)}{\alpha}, a_1 = -\frac{12\sqrt{-(h-2k)\alpha}\,(-4k+h(2+\alpha\epsilon))^{3/2}\mu}{\alpha^2(-6k+h(3+\alpha\epsilon))},$$

$$a_2 = \frac{12(-4k+h(2+\alpha\epsilon))^2 \mu^2}{\alpha^2(-6k+h(3+\alpha\epsilon))}, \lambda = -\frac{\sqrt{-(h-2k)\alpha}}{\sqrt{2h-4k+h\alpha\epsilon}}, \beta = -2; \qquad (10)$$

Putting Eq. (10) into Eq. (9), the new wave soliton solution of (GPE) can be obtained as

$$u_1(x,t) = -\frac{2(h-2k)}{\alpha} + \frac{12(-4k+h(2+\alpha\epsilon))^2\mu^2}{\alpha^2(-6k+h(3+\alpha\epsilon))\left(-\frac{L\mu}{M} + C_1\cosh\left(\frac{M\xi}{L}\right) - C_1\sinh\left(\frac{M\xi}{L}\right)\right)^2}$$

$$+ \frac{12M(-4k+h(2+\alpha\epsilon))^{3/2}\mu}{\alpha^2(-6k+h(3+\alpha\epsilon))\left(-\frac{L\mu}{M} + C_1\cosh\left(\frac{M\xi}{L}\right) - C_1\sinh\left(\frac{M\xi}{L}\right)\right)}, \qquad (11)$$

where $M = \sqrt{(-h+2k)\alpha}, L = \sqrt{2h-4k+h\alpha\epsilon}$.

Substituting the suitable values $h=2$, $k=2.5$, $\alpha=2.6$, $\mu=0.2$, $\epsilon=3.5$, $C_1=0.6$ into Eq. (11), 3D, 2D, contour surface of Eq. (1) can be seen in Fig. 1.

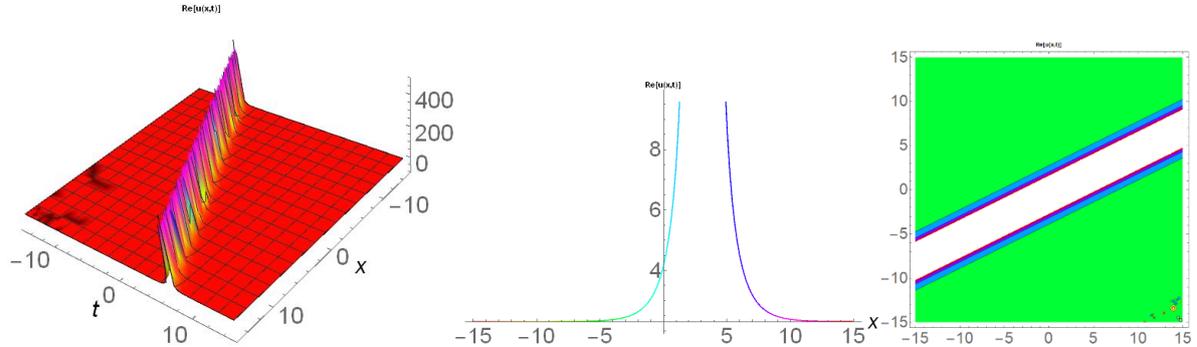

Fig.1. 2D and 3D surfaces of Eq. (11)

Set 2:
$$a_0 = 0, a_1 = \frac{12h^{3/2}\sqrt{h-2k}\epsilon^{3/2}\mu}{2k+h(-1+\alpha\epsilon)}, a_2 = \frac{12h^2\epsilon^2\mu^2}{2k+h(-1+\alpha\epsilon)}, \lambda = \frac{\sqrt{h-2k}}{\sqrt{h}\sqrt{\epsilon}}, \beta = -2; \quad (12)$$

If these are getting into Eq. (9), subsequently, another new traveling wave solutions of the (GPE) can get as

$$u_2(x,t) = \frac{12h^2\epsilon^2\mu^2}{\left(2k+h(-1+\alpha\epsilon)\right)\left(-\frac{\sqrt{h}\sqrt{\epsilon}\mu}{\sqrt{h-2k}}+C_1\cosh\left(\frac{\sqrt{h-2k}\xi}{\sqrt{h}\sqrt{\epsilon}}\right)-C_1\sinh\left(\frac{\sqrt{h-2k}\xi}{\sqrt{h}\sqrt{\epsilon}}\right)\right)^2} + \frac{12h^{3/2}\sqrt{h-2k}\epsilon^{3/2}\mu}{\left(2k+h(-1+\alpha\epsilon)\right)\left(-\frac{\sqrt{h}\sqrt{\epsilon}\mu}{\sqrt{h-2k}}+C_1\cosh\left(\frac{\sqrt{h-2k}\xi}{\sqrt{h}\sqrt{\epsilon}}\right)-C_1\sinh\left(\frac{\sqrt{h-2k}\xi}{\sqrt{h}\sqrt{\epsilon}}\right)\right)}. \quad (13)$$

Putting suitable values $h=4$, $k=0.5$, $\alpha=2.6$, $\mu=0.2$, $\epsilon=4$, $C_1=0.3$ into Eq. (13), 3D, 2D, contour surface of Eq. (1) can appear in Fig. 2.

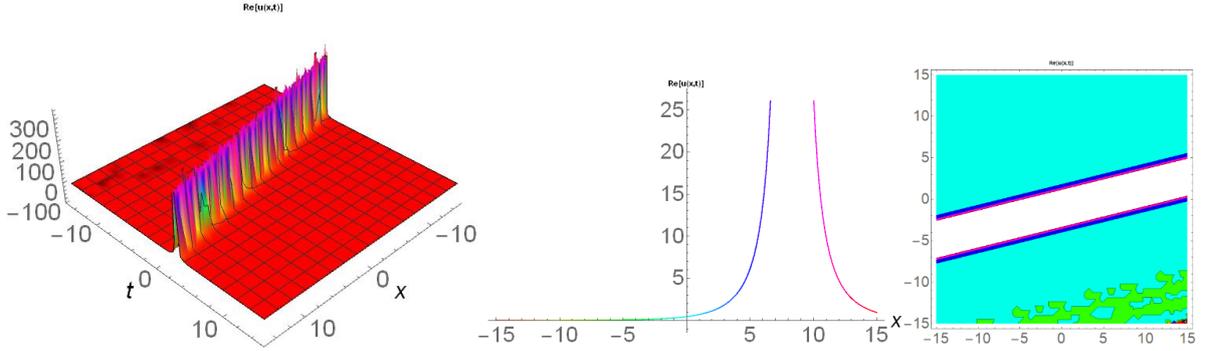

Fig.2. 2D and 3D surfaces of Eq. (13).

Set 3:
$$a_0 = \frac{4k\epsilon\lambda^2}{\alpha+(2+\alpha\epsilon)\lambda^2}, a_2 = \frac{(\alpha-\lambda^2)(\alpha+(2+\alpha\epsilon)\lambda^2)a_1^2}{24k\alpha\epsilon\lambda^2},$$
$$\mu = \frac{(\alpha-\lambda^2)(\alpha+(2+\alpha\epsilon)\lambda^2)a_1}{24k\alpha\epsilon\lambda}, \beta = -2, h = \frac{2k(\alpha+2\lambda^2)}{\alpha+(2+\alpha\epsilon)\lambda^2}; \quad (14)$$

These give another novel exact solution of (GPE) as bellow

$$u_3(x,t) = \frac{4k\epsilon\lambda^2}{\alpha+(2+\alpha\epsilon)\lambda^2} + \frac{\left(\alpha-\lambda^2\right)\left(\alpha+(2+\alpha\epsilon)\lambda^2\right)a_1^2}{24k\alpha\epsilon\lambda^2\left(C_1\cosh(\lambda\xi)-C_1\sinh(\lambda\xi)-\dfrac{\left(\alpha-\lambda^2\right)\left(\alpha+(2+\alpha\epsilon)\lambda^2\right)a_1}{24k\alpha\epsilon\lambda^2}\right)^2}$$

$$+ \frac{a_1}{C_1\cosh(\lambda\xi)-C_1\sinh(\lambda\xi)-\dfrac{\left(\alpha-\lambda^2\right)\left(\alpha+(2+\alpha\epsilon)\lambda^2\right)a_1}{24k\alpha\epsilon\lambda^2}}.$$

(15)

Putting the values $k=2$, $\alpha=5$, $\lambda=1.2$, $\epsilon=6.6$, $C_1=2, a_1=2.8$ into Eq. (15), 3D, 2D, contour surface of Eq. (1) can be shown in Fig. 3.

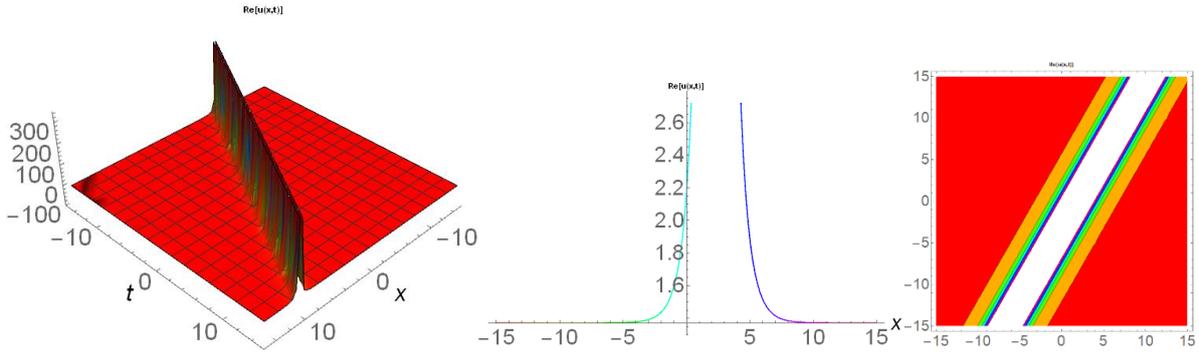

Fig.3. 2D and 3D surfaces of Eq. (15).

Set 4:

$$a_0 = \frac{4k\epsilon}{1+\alpha\epsilon},\ a_2 = 0,\ \beta = -3,\ \mu = \frac{i\sqrt{\alpha}(1+\alpha\epsilon)a_1}{4k\epsilon},\ h = \frac{2k}{1+\alpha\epsilon},\ \lambda = i\sqrt{\alpha};$$

(16)

Using these values into Eq. (9), the new complex solutions of the (GPE) can be produced as

$$u_4(x,t) = \frac{4k\epsilon}{1+\alpha\epsilon} + \frac{a_1}{C_1\cos(\sqrt{\alpha}\xi)-iC_1\sin(\sqrt{\alpha}\xi)-\dfrac{(1+\alpha\epsilon)a_1}{4k\epsilon}}.$$

(17)

Putting suitable values $k=4.5$, $\alpha=0.4$, $\epsilon=0.3$, $C_1=0.2, a_1=0.8$ into Eq. (17), 3D, 2D, contour surface of Eq. (1) can be shown in Fig. 4.

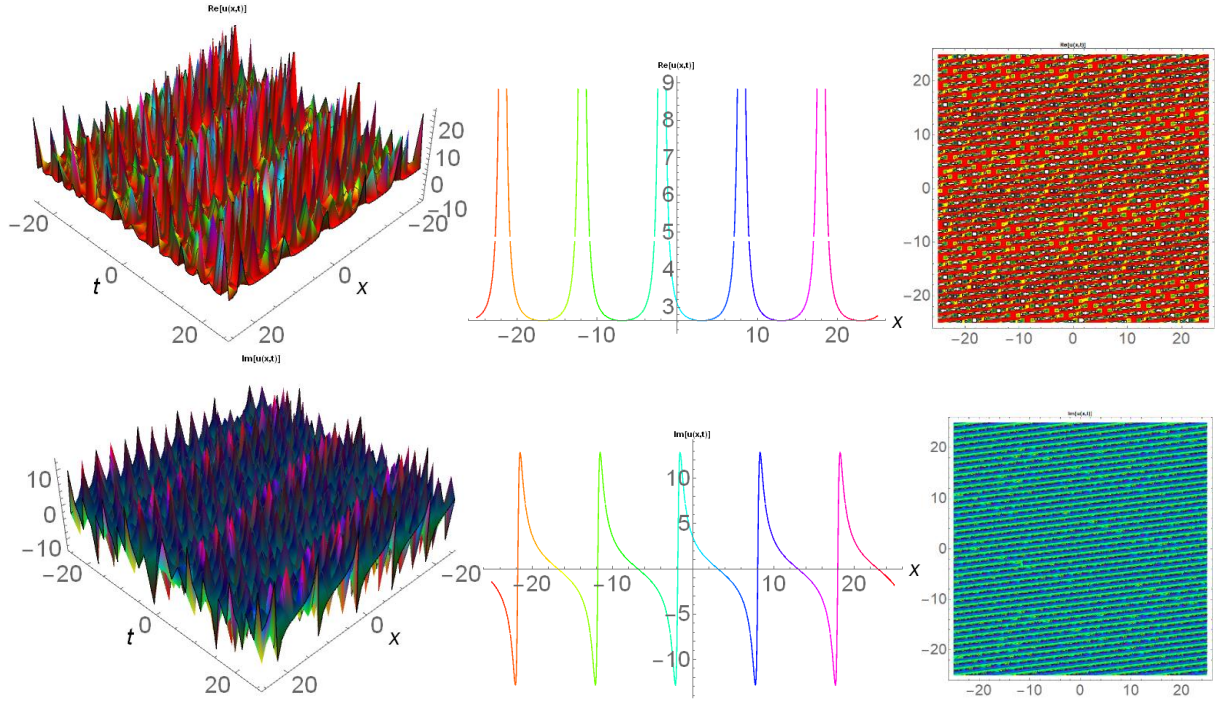

Fig.4. 2D and 3D surfaces of Eq.(17).

Set 5:

$$a_0 = \frac{i\sqrt{\alpha}a_1}{\mu},\ a_2 = 0,\ \beta = -3,\ h = \frac{i\sqrt{\alpha}a_1}{2\epsilon\mu},\ k = \frac{i\sqrt{\alpha}(1+\alpha\epsilon)a_1}{4\epsilon\mu},\ \lambda = i\sqrt{\alpha};$$
(18)

Considering Eq. (18) into Eq. (9) produces another novel complex solution of (GPE) as follows

$$u_5(x,t) = \frac{i\sqrt{\alpha}a_1}{\mu} + \frac{a_1}{\frac{i\mu}{\sqrt{\alpha}} + C_1 \cos(\sqrt{\alpha}\xi) - iC_1 \sin(\sqrt{\alpha}\xi)}.$$
(19)

Putting the values $\mu = 0.4,\ \alpha = 0.4,\ \epsilon = 0.3,\ C_1 = 0.2, a_1 = 0.8$ into Eq. (19), 3D, 2D, contour surface of Eq. (1) can be shown in Fig. 5.

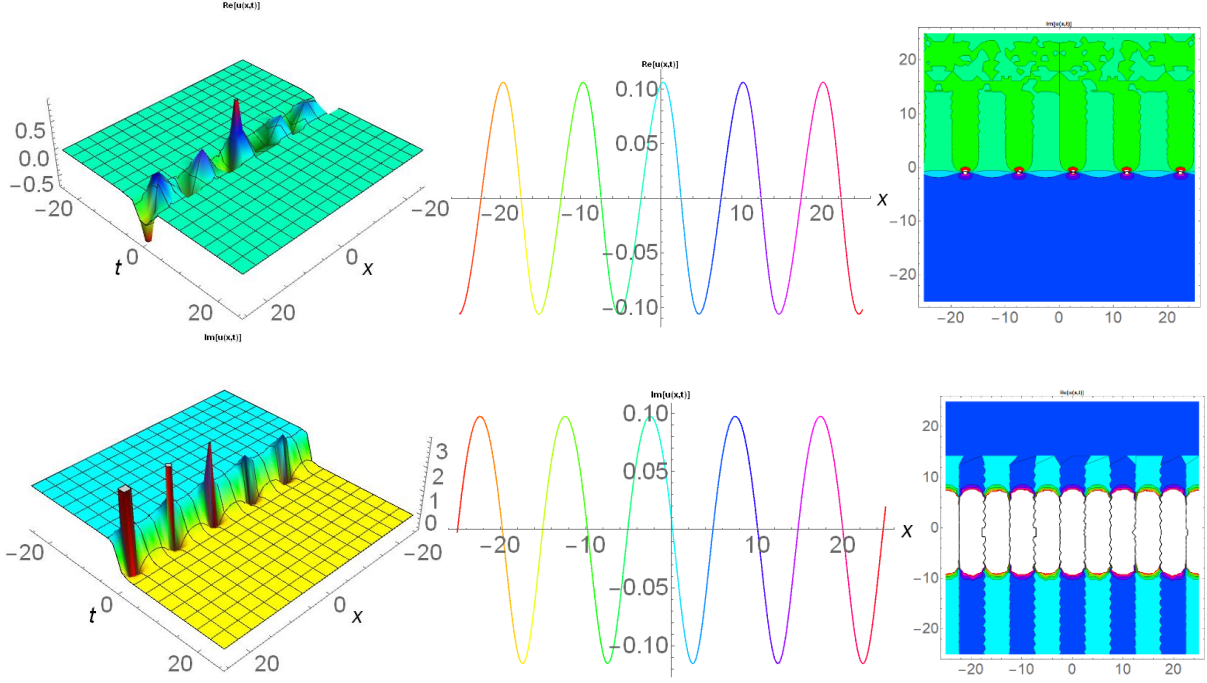

Fig.5. 2D and 3D surfaces of Eq. (19).

Set 6:

$$a_0 = \frac{12h\epsilon\mu + 3\lambda a_1 - \sqrt{-96h\epsilon\lambda\mu a_1 + 9(4h\epsilon\mu + \lambda a_1)^2}}{24\mu}, \ a_2 = \frac{\mu a_1}{\lambda}, \ \beta = -2,$$

$$\alpha = \frac{\lambda\left(12h\epsilon\mu - \lambda a_1 + \sqrt{3}\sqrt{48h^2\epsilon^2\mu^2 + \lambda a_1(-8h\epsilon\mu + 3\lambda a_1)}\right)}{2a_1},$$

$$k = \frac{24h\mu + 12h\epsilon\lambda^2\mu - 3\lambda^3 a_1 + \lambda^2\sqrt{-96h\epsilon\lambda\mu a_1 + 9(4h\epsilon\mu + \lambda a_1)^2}}{48\mu};$$

(20)

Using Eq. (20) into Eq. (9), another new soliton solution of (GPE) can be derived as

$$u_6(x,t) = \frac{\mu a_1}{\lambda\left(-\frac{\mu}{\lambda} + C_1\cosh(\lambda\xi) - C_1\sinh(\lambda\xi)\right)^2} + \frac{a_1}{-\frac{\mu}{\lambda} + C_1\cosh(\lambda\xi) - C_1\sinh(\lambda\xi)} +$$

$$\frac{12h\epsilon\mu + 3\lambda a_1 - \sqrt{-96h\epsilon\lambda\mu a_1 + 9(4h\epsilon\mu + \lambda a_1)^2}}{24\mu}.$$

(21)

Substituting the values $\mu = 1.5, \ \alpha = 0.4, \ \epsilon = 0.1, \ C_1 = 2, a_1 = 0.4, \ h = -1, \ \lambda = 0.5$ into Eq. (21), 3D, 2D, contour surface of Eq. (1) can be shown in Fig. 6.

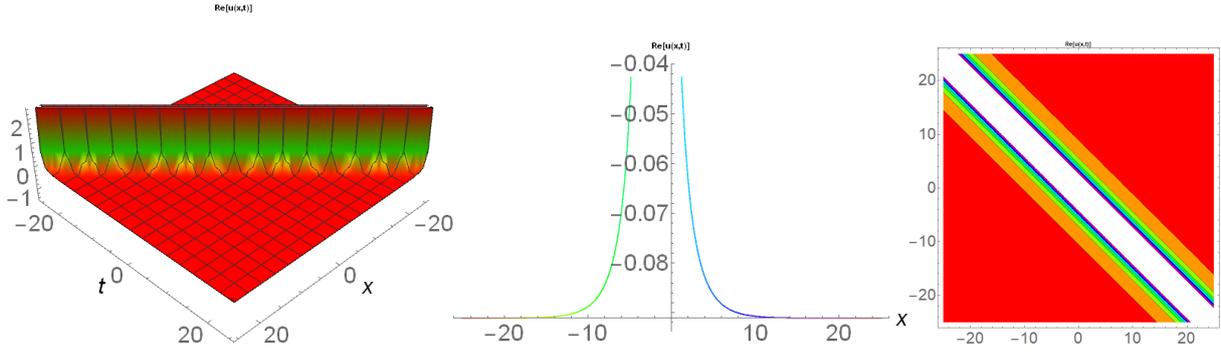

Fig.6. 2D and 3D surfaces of Eq. (21).

**Set 7:**

$$a_0 = \frac{i\sqrt{\alpha}\, a_1}{\mu},\ a_2 = 0,\ \beta = -3,\ k = \frac{h}{2} + \frac{i\alpha^{3/2} a_1}{4\mu},\ \lambda = i\sqrt{\alpha},\ \epsilon \frac{i\sqrt{\alpha}\, a_1}{2h\mu}; \quad (22)$$

Eq. (22) gives

$$u_7(x,t) = \frac{i\sqrt{\alpha}\, a_1}{\mu} + \frac{a_1}{\dfrac{i\mu}{\sqrt{\alpha}} + C_1 \cos(\sqrt{\alpha}\,\xi) - iC_1 \sin(\sqrt{\alpha}\,\xi)}. \quad (23)$$

Putting the values $\mu = 2.5,\ \alpha = 4,\ C_1 = 2.5,\ a_1 = 0.2,\ h = 2$ into Eq. (23), 3D, 2D, contour surface of Eq. (1) can be seen in Fig. 7.

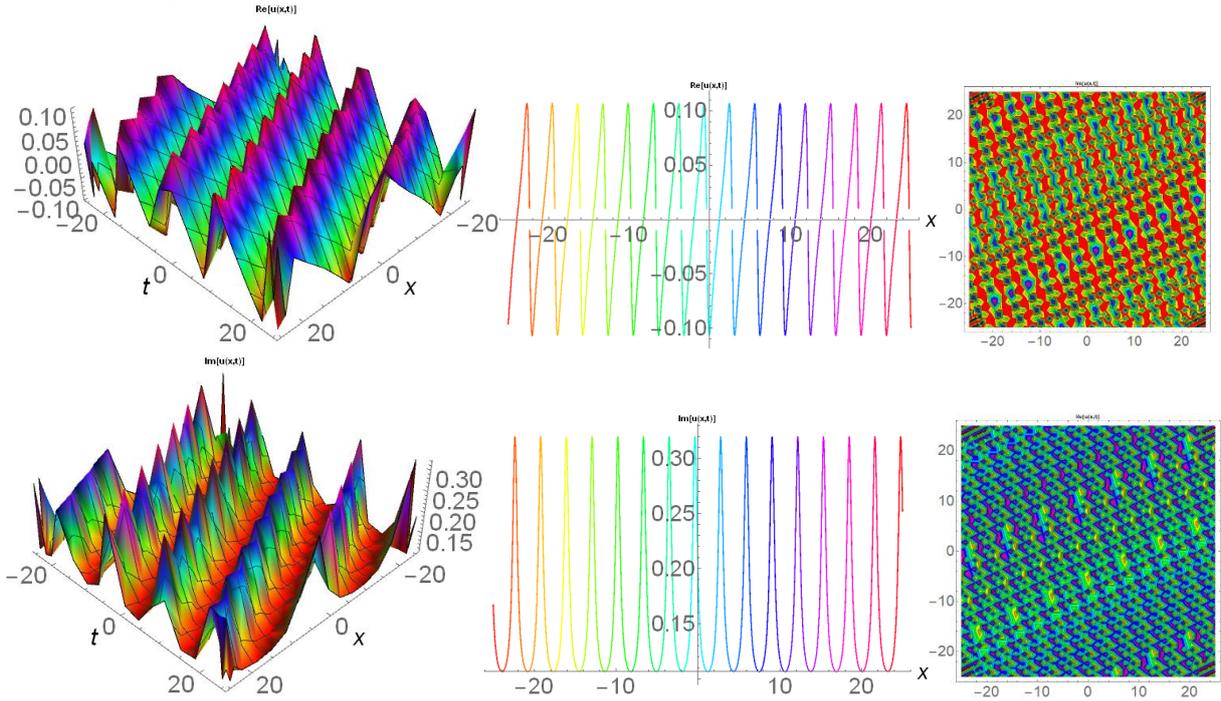

Fig.7. 2D and 3D surfaces of Eq. (23).

**Set 8:**

$$a_0 = -\frac{2(h-2k)}{\alpha},\ a_2 = 0,\ \beta = -3,\ \mu = \frac{i\alpha^{3/2} a_1}{2(h-2k)},\ \lambda = -i\sqrt{\alpha},\ \epsilon \frac{h-2k}{h\alpha}; \quad (24)$$

Eq. (24) gives another new complex solution of (GPE) as below

$$u_8(x,t) = -\frac{2(h-2k)}{\alpha} + \frac{a_1}{C_1 \cos(\sqrt{\alpha}\xi) + iC_1 \sin(\sqrt{\alpha}\xi) + \frac{\alpha a_1}{2(h-2k)}}. \tag{25}$$

Putting the values $\mu = 0.5$, $\alpha = 4$, $C_1 = 2.5$, $a_1 = 0.2$, $h = 0.2$, $k = 0.2$ into Eq. (25), 3D, 2D, contour surface of Eq. (1) can seem in Fig. 8.

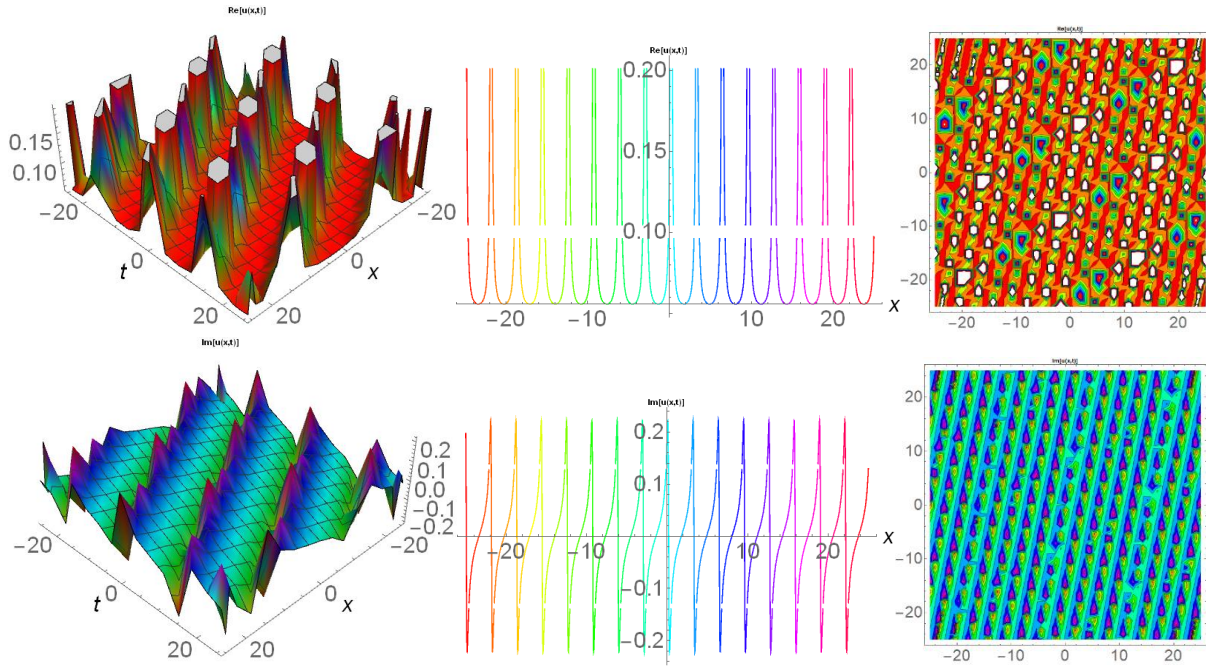

Fig.8. 2D and 3D surfaces of Eq. (25).

## 4. Conclusion

In this study, we have been successfully applied the $(1/G')$-expansion method to find new exact analytical solutions of the Gilson-pickering equation. The solution that we obtain are completely new solutions which conclude of complex solutions, trigonometric solution, and singular solutions, it is due to the efficiency of the applied method. For better understanding the physical structure of the produced solutions, we plotted the 2D, 3D dimensional and contour surface with using the suitable values of the parameters.